\documentclass[apjl,iop]{emulateapj}

\shorttitle{Determination of mass of IGR J17091-3624}
\shortauthors{Iyer et al.}

\usepackage{graphicx}
\usepackage{natbib}

\newcommand{\Msun}{\,M_{\odot}}
\newcommand{\swift}{\it Swift}
\newcommand{\xte}{\it RXTE}
\newcommand{\intg}{\it INTEGRAL}

\begin{document}

\title{Determination of mass of IGR J17091-3624 from ``spectro-temporal'' variations during
onset-phase of the 2011 outburst}

\author{N.~Iyer\altaffilmark{1,2} and A.~Nandi\altaffilmark{1}} 
\affil{$^1$Space Astronomy Group, SSIF/ISITE Campus, ISRO Satellite Centre, Outer Ring Road,
Marathahalli, Bangalore, 560037, India}
\affil{$^2$Indian Institute of Science, Bangalore, 560012, India}

\and

\author{S.~Mandal\altaffilmark{3}}
\affil{$^3$Indian Institute of Space Science and Technology, Trivandrum, 695547, India}
\email{samir@iist.ac.in}

\begin{abstract}
The 2011 outburst of the black hole candidate IGR J17091-3624 followed the canonical track of 
state transitions along with the evolution of Quasi-Periodic Oscillation (QPO) frequencies before 
it began exhibiting various variability classes similar to GRS 1915+105. We use this canonical 
evolution of spectral and temporal properties to determine the mass of IGR J17091-3624, 
using three different methods, viz: {\it Photon Index ($\Gamma$) - QPO frequency ($\nu$) correlation},
{\it QPO frequency ($\nu$) - Time ($day$) evolution} and {\it broadband spectral modelling based 
on Two Component Advective Flow}. We provide a combined mass estimate for 
the source using a Naive Bayes based joint likelihood approach. This gives
a probable mass range of ${\rm 11.8\Msun - 13.7\Msun}$. Considering
each individual estimate and taking the lowermost and uppermost bounds among all the three methods,
we get a mass range of $8.7\Msun - 15.6\Msun$ with 90\% confidence.
We discuss the possible implications of our 
findings in the context of two component accretion flow.
\end{abstract}


\section{Introduction}\label{s:int}

The enigmatic Galactic black hole candidate IGR J17091-3624 underwent
multiple outbursts in 1994, 2001, 2003 and 2007 before the recent one in February 2011 
\citep{int03,rev03,cap06,cap12}. During the 2011 outburst, the source 
displayed state transitions like other outbursting black hole sources \citep{rem06,nan12} from the Hard to 
Intermediate states \citep{pah11,rod11,cap12,iye13}. It also showed presence of a 
state with characteristics similar to a `canonical' soft state 
\citep{iye13} of other outbursting black hole sources. After evolving along the 
canonical track, it started to display a number of
temporal variability classes similar to GRS 1915+105 \citep{alt11}. 
The presence of High Frequency QPO ($\sim$ 66 Hz)
similar to GRS 1915+105 \citep{alt12} and radio detections, albeit at the level of a few mJy 
\citep{rod11} in the Hard and the Intermediate states were also seen in this outburst.

The similarity to GRS 1915+105 triggered immense interest about
the nature of this source. As of now, its precise location upto $5{\arcsec}$ 
is known and possible optical and infrared counterparts have been identified \citep{bod12} 
within the {\it Chandra} error circle.
The distance to the source, mass and orbital parameters of both objects 
in the binary are currently unknown. What is certainly known, 
is that IGR J17091-3624 is only the second black hole candidate to display a wide range of
temporal and spectral variations 
after GRS 1915+105, although at much lower (10 to 50 times) observed flux levels \citep{alt11}.
More knowledge about IGR J17091-3624 should 
allow us then to make some meaningful comparisons between these two sources.  
Multiple attempts have been made before to determine the mass of IGR J17091-3624. These attempts
have placed it anywhere between $\mathrm{< 3\Msun}$ 
\citep[]{alt11} to 
$\mathrm{ < 15\Msun}$ \citep[]{alt12}. Other attempts to estimate the mass are 
$\mathrm{ < 5\Msun}$ \citep[]{rao12}, $\mathrm{ 9.13 \pm 2.25\Msun}$ 
\citep[]{pah11}, and $\mathrm{ \sim 6\Msun}$ \citep[]{reb12}. 
The wide range of these estimates doesn't enable us to even pin down whether the source
is a massive stellar black hole binary or an exceptionally low mass black hole candidate.

In this paper, we attempt to make an estimate of the mass with tighter constraints
using X-ray observations made during the rising phase of its 2011 outburst. Here, we discuss
{\it three} different methods (two of which are independent of distance to the source) 
to estimate its mass. In the 3$^{rd}$ approach, we study the broadband (0.5 to 100.0 keV)
spectrum using the two component advective flow (TCAF) model \citep{cha95,cha06}.
We use these three estimates to put a limit on the mass of the central object. 
Finally, we also discuss the 
results obtained from our spectro-temporal analysis in the light of two different types of 
accretion flows i.e., a Keplerian disk \citep{sha73} on the equatorial plane, 
sandwiched by a sub-Keplerian flow \citep{cha95,cha06,wu02,smi01,smi02,smi07,cam13} 
on both sides of the Keplerian disk in the vicinity of an accreting black hole.

\section{Observations and Analysis}\label{s:obs}

We use archival data of observations made by the {\it RXTE}, {\it Swift} and {\it INTEGRAL} 
satellites for the 2011 outburst (from \texttt{06 Feb (MJD 55598.28)} onwards). We restrict ourselves to the rising 
phase of the outburst, before the source entered into its enigmatic and unpredictable variability 
phase. In all, we analyze $\sim$ 40 days of data from these observatories. 

\subsection{\swift{} data reduction}
We use the {\it XRT} data from the {\it Swift} suite of instruments. 
The {\it XRT} data consists of windowed timing mode observations, which reduces
pile-up effects from the source.  The maximum observed count rate is 
about $40{\rm~counts~s^{-1}}$. This ensures that the XRT data is not piled up \citep{rom06}. 
We analyse these {\it XRT} data-sets by following the steps as given in the XRT manual\footnote{Swift XRT Data Reduction Guide, v1.2, 2005}.
For this purpose, we use \texttt{HEASoft 6.15.1} and its associated ftools packages.
We restrict ourselves to use events of {\it grade 0-2}, while extracting timing and
spectral data. The source and background regions are extracted by selecting a circular region of 
diameter 40 pixels ($\sim{\rm 90\arcsec}$) from the data images.
While doing spectral modelling, we found that the low energy spectrum 
(0.5 -- 1.0 keV) of the {\it Swift XRT} data did not fit well and showed 
some excess in the soft state. This was due to uncertainty in position of the source in the 
binned pixels of WT mode\footnote{http://www.swift.ac.uk/analysis/xrt/digest\_cal.php\#abs}.
By using position dependent {\it rmfs}\footnote{http://www.swift.ac.uk/analysis/xrt/rmfs.php}
made for different probable positions of the source on the detector Y pixel,
we accounted for this apparent excess in the soft spectrum 
(K. Page 2014, private communication).
We use such position dependent {\it rmfs} for each of our {\it Swift XRT} data-sets.

\subsection{\intg{} data reduction}
We use the {\it IBIS/ISGRI} instrument \citep{leb03} from the {\it INTEGRAL} suite.
We extract spectral information from {\it ISGRI } by
following the steps as given in their data reduction manual\footnote{IBIS Analysis User Manual, Chernyakova, M. et~al., 2012, ISDC}.
To obtain the spectrum, we use {\tt OSA v 10.0} along with updated versions
of the calibration and response files. The spectrum from {\it ISGRI} is
rebinned in order to get more data points for modelling the spectrum (see
  Table \ref{tab:sys}).
We extract spectral information from {\it ISGRI} only for those dates 
which have simultaneous {\it Swift} observations. This helps us to obtain 
broadband (0.5 -- 100 keV) data for our spectral analysis. 
In all the data-sets that we use, IGR J17091-3624 is visible in the
{\it ISGRI} detector with significance greater than 7$\sigma$.
However, owing to the low source flux, there is non-detection or very
low significance detection in the {\it JEM-X} detector. Hence, 
we do not use any spectral data from {\it JEM-X}.

\subsection{ \xte{} data reduction}
We use the {\it PCU2} and {\it HEXTE} detector data of the {\it RXTE} satellite. 
However, as noted \citep{pah11,cap12,iye13}, the observations 
made by {\it RXTE} before \texttt{23 Feb} are contaminated due to the nearby 
LMXB NS source GX 349+2. 
We do use these contaminated observations for our timing analysis.
Specifically, we use them to find the frequency of QPOs from the power density 
spectrum (PDS). The source GX 349+2 does not exhibit any QPO like features in its PDS.
We investigated the observations carried out by the {\it JEM-X} payload onboard {\it INTEGRAL} 
on \texttt{11 Feb 2011} to ascertain this. Also, as noted by \citet{one02,agr03} no 
QPOs have ever been found in GX 349+2, in-spite of extensive searches carried 
out over all the spectral and temporal states of this LMXB. This leads us to
believe that the QPO like features in the PDS obtained from {\it PCU2} data 
are indeed from IGR J17091-3624. We use only the non contaminated data 
from {\it PCU2} and {\it HEXTE} detector for our spectral analysis.

\subsection{Temporal Analysis}\label{ss:tem}

The Power Density Spectrum (PDS) of all data-sets is computed in
the energy bands afforded by the observing instruments, viz. {\it Swift XRT} (0.5 -- 10.0 keV)
and {\it RXTE PCA} (3.0 -- 30.0 keV). The PDS are obtained using \texttt{GHATS v1.1}\footnote{http://www.brera.inaf.it/utenti/belloni/GHATS\_Package/Home.html}, 
a customized IDL based timing package, which takes care of re-binning, Poisson noise subtraction and dead-time 
correction \citep{zha95} of the FFT data. Each of the PDS is made by using temporal data of bin size 3.52 ms
(for {\it Swift XRT} data) and 0.244 ms (for {\it RXTE PCA} data). All observations are divided
into segments of 128.0 s and a PDS is computed individually for each such segment. We then
obtain the final PDS for an entire observation by averaging the PDS from each individual
segment. Finally, we systematically search each averaged PDS for the presence of QPOs.
A combination of Lorentzian features is used to model the PDS.
We select only those QPO features which have a significance greater than $3 \sigma$ for our analysis. 
The Q-factor ($\nu / FWHM$) of these selected QPOs is found to vary between 3 to 10.
The rms power under the QPO varies in between 8\% to 17\% and
the QPO frequencies are observed to increase with time from 0.055 Hz to $\sim $ 5 Hz
\citep{sha11,pah11,rod11,iye13}.  
We have modelled this evolution of QPO frequencies in \S \ref{ss:qpo}.
The error on QPO frequency is taken to be 
1$\sigma$ deviation from its centroid frequency, which is obtained by scaling its FWHM. 

\subsection{Spectral Analysis}

The spectral analysis is carried out using \texttt{XSPEC v12.8.1g}.
We do spectral modelling for individual observations of each instrument.
Apart from this we also analyse broadband 
spectra (0.5 keV to 100.0 keV) as obtained by choosing {\it simultaneous observations} 
from {\it Swift XRT} and {\it INTEGRAL IBIS}. 
We rebin the spectrum obtained from each instrument and add systematic errors
as mentioned in Table \ref{tab:sys} before fitting the data-sets.
We model the energy spectrum by two different methods. The first method 
\citep[see][]{cap12,iye13} is a phenomenological model fitting of
the spectrum and consists of a \texttt{diskbb} and a \texttt{powerlaw} (or \texttt{cutoffpl}) 
component.  
In this phenomenological modelling, we find that the \texttt{powerlaw} photon index ($\Gamma$)          
increases with time and is positively correlated with the QPO frequencies obtained in \S\ref{ss:tem}.
We model this correlation as presented in \S\ref{ss:qnu} to obtain the source mass.
We use 1$\sigma$ errors on fit parameters as estimated from the \texttt{XSPEC
error} command for our model fitting routines.
In the second method of spectral modelling, we calculate a model spectrum for each 
broadband observation using Two Component Advective Flow (TCAF).
This is done by including radiative hydrodynamics of the accreting plasma self-consistently 
into the governing equations of the flow \citep{cha95,cha06}. We then use this calculated
spectrum to fit the observed spectrum, which is presented in \S\ref{ss:spe}.

\begin{deluxetable*}{ccc}
  \tabletypesize{\small}
  \tablecaption{Reconditioning of reduced spectral data-sets \label{tab:sys}}
  \tablewidth{0pt}
  \tablehead{
	\colhead{Instrument} & \colhead{Binning} & \colhead{Systematic error}
  }
  \startdata
  \swift{} XRT & Minimum 20 counts per bin & 3\% \\
  \intg{} IBIS/ISGRI & 25 bins between 13 keV -- 150 keV$^{a}$ & 5\% \\
  \xte{} PCA & None & 0.5\% \\
  \enddata
  \tablenotetext{a}{IBIS/ISGRI has Gaussian distributed errors and doesn't need rebinning for noise statistics. 
	By default 11 bins are present between 13 keV -- 150 keV.}
\end{deluxetable*}

\section{Models and Results}\label{s:res}

We consider three different approaches to estimate mass of the central source. These methods are 
discussed in the following sub-sections: spectro-temporal correlation (\S \ref{ss:qnu}), 
the evolution of QPO frequencies (\S \ref{ss:qpo}) and broadband spectral modelling (\S \ref{ss:spe}). 
Then, we use these three different estimates to derive a single set of bounds for the mass. 
The estimates are made under the TCAF paradigm \citep[]{cha95}, 
which enables us to model both evolution of spectral and temporal features under a single paradigm, 
thereby ensuring consistency of the estimate. In this paper, we have used radial distance in units of 
Schwarzschild radius ($r_g$) and mass in units of solar mass ($M_\odot$).

\subsection{Photon Index ($\Gamma$) - QPO frequency ($\nu$) correlation} \label{ss:qnu}

A model for the observed correlation between the spectral fitted photon index ($\Gamma$) 
and the observed QPO frequency ($\nu$) for black hole candidates is presented 
in \citet{tit04}. \citet{sha07} (hereafter ST07) have presented an implicit empirical
scaling relation between the correlation curves of different black hole sources and used 
it to estimate the mass of a few such sources. The model uses a `Compton cloud' around
the central source and a transition layer / region between this Compton cloud and the
Keplerian disk. Any change in the size of the cloud leads to a change in both the QPO 
frequency (explained as the magneto-acoustic resonance oscillations of 
the bounded transition layer \citep{tit02}) and the photon index of the emitted spectrum 
($\Gamma$) (due to varying optical depth of the hot electrons). We believe that the 
`Compton cloud' in this model is the same as the post-shock region of the 
accretion disk under 
the TCAF paradigm \citep{cha95}. The TCAF paradigm too demonstrates both the
steepening of spectral index and the increase in QPO frequencies with time \citep{cha08,nan12,rad14,deb14}.
In addition, the accretion disk configuration used in
\citet{tit02,tit04} is similar to that under the TCAF paradigm. 
We thus expect the empirical correlation of ST07 should hold under the TCAF paradigm as well.

The  correlation is fitted using the empirical relation given in ST07 :

\begin{equation}
  \Gamma(\nu) = A - DB\ln\left[exp\left(\frac{\nu_{tr} - \nu}{D}\right) + 1\right],
  \label{eq:gnc}
\end{equation}

where A is the value at which the photon index saturates, $\nu_{tr}$ is the threshold 
frequency above which the levelling off/saturation of $\Gamma$ happens and B is the slope of 
the correlation which scales with mass. The parameter D controls how fast (i.e., over what 
frequency range) the transition occurs. 
We note that this is an empirical scaling equation relating the shape of the correlation curve with the
masses of different black hole systems. To find the mass of an unknown source requires comparison of its
correlation curve with a reference curve of a source with known mass. An inherent assumption in this scaling equation
is the similarity of the shape of the correlation curves between these two sources \citep[see also][]{sha09}.
The slope of this correlation curve, as the source transits from low photon index and frequency values in 
the harder states to the higher saturated photon index and frequency values of the softer states,
is related to the mass of the source. The empirical scaling relation estimates this slope and a 
comparison of slopes from different sources gives a direct comparison of their masses.
\begin{figure}
\epsscale{1.1}
\plotone{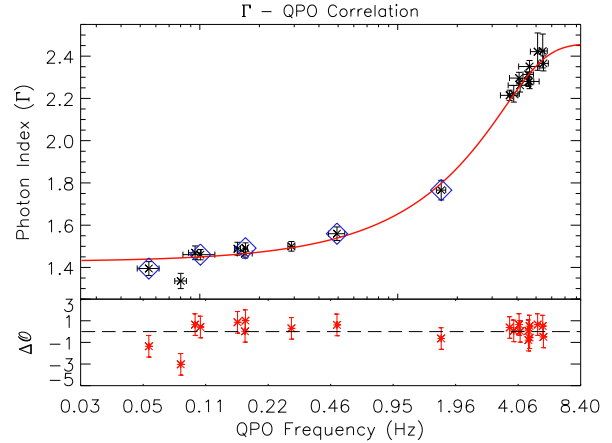}
\caption{Photon index ($\Gamma$) - QPO frequency ($\nu$) correlation plot. 
Diamond points (blue) are {\it Swift XRT} QPOs. Rest of the 
QPO frequencies are from {\it RXTE PCA}. Bottom panel shows the weighted residuals in each point.
The weights are assigned as given in Appendix I.
The best fit curve (red solid line) using 
Eqn. \ref{eq:gnc} gives mass to be $\mathrm{10.90^{-1.48}_{+1.67}\Msun}$.}
  \label{fig:gnc}
\end{figure}
We consider the rising phase of the 2005 outburst of GRO~J1655-40
as the reference correlation curve for estimating mass of IGR J17091-3624. 
This correlation was successfully used in ST07 to predict
the mass of GRS 1915+105, a source with a very similar correlation curve to IGR J17091-3624.
We keep the value of D fixed at 1.0 as done in ST07.
We note though, that changing the 
value of D changes the final value of mass by a small amount.
The best fitted curve (for D = 1.0)  using Eqn. \ref{eq:gnc} is shown in Fig. \ref{fig:gnc} along-with the 
residuals (bottom panel). This fitting has been done using Craig Markwardt's IDL based routines\footnote{http://purl.com/net/mpfit \label{fn:mrk}} \citep{mar09}, suitably modified for 
accommodating errors in both variables, as is explained further in Appendix I. 
This fitting gives us values as A = $2.38^{-0.05}_{+0.09}$, 
B = $0.23^{-0.02}_{+0.03}$ and $\nu_{tr} = 4.47^{-0.64}_{+0.87}$ Hz.  
The best fit mass (as obtained by scaling the B parameter with GRO J1655-40's mass which
is taken to be $6.3 \pm 0.5\Msun$ \citep{grn01})
comes as $\mathrm{M_{bh} = 10.90^{-1.48}_{+1.67}\Msun}$ with a fit statistic (Appendix I) of
$\mathcal{O} = 16.964$ for 18 degrees of freedom.  

\subsection{QPO frequency ($\nu$) - Time ($day$) evolution} \label{ss:qpo}

\begin{figure}[h]
\epsscale{1.8}
\plottwo{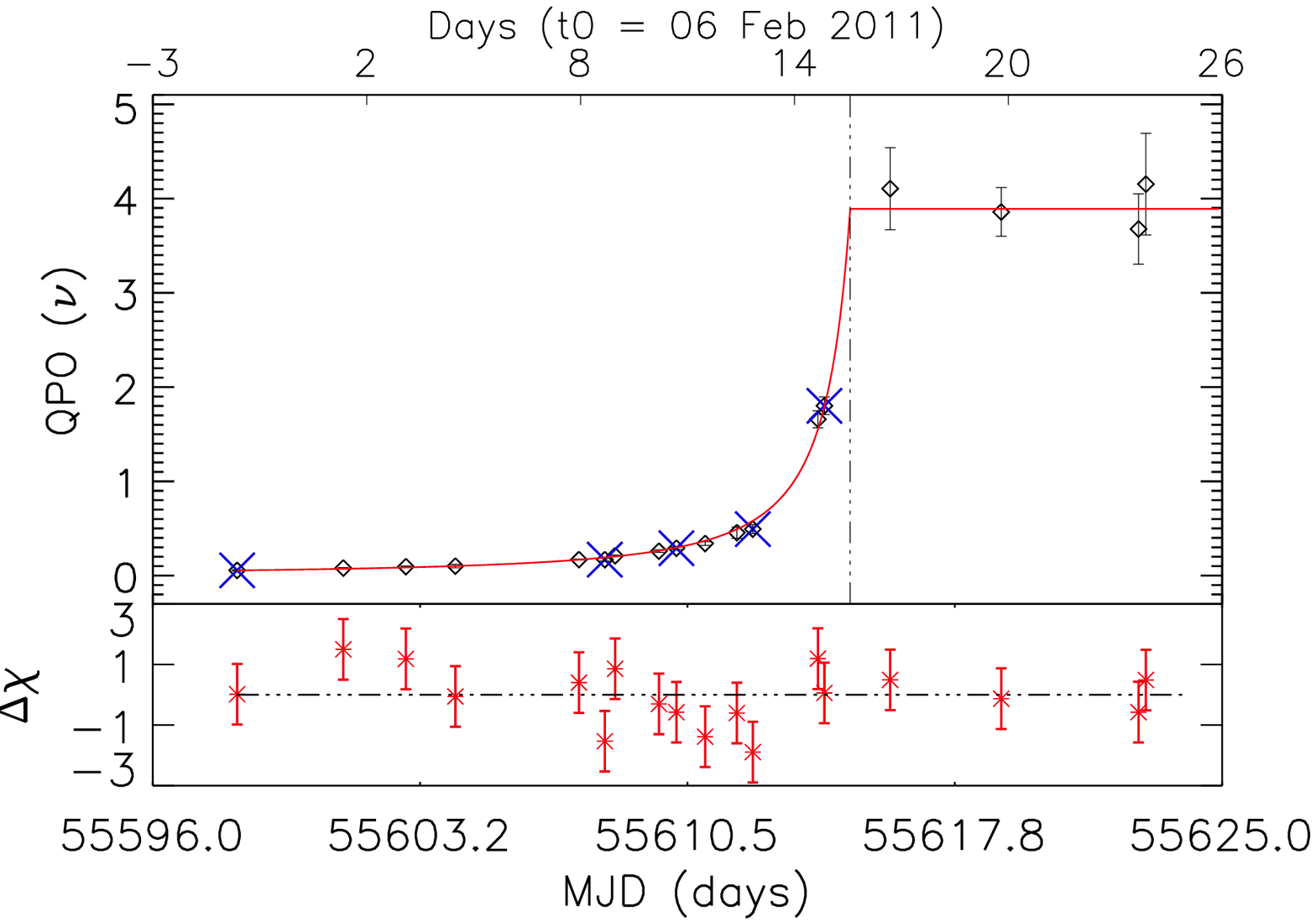}{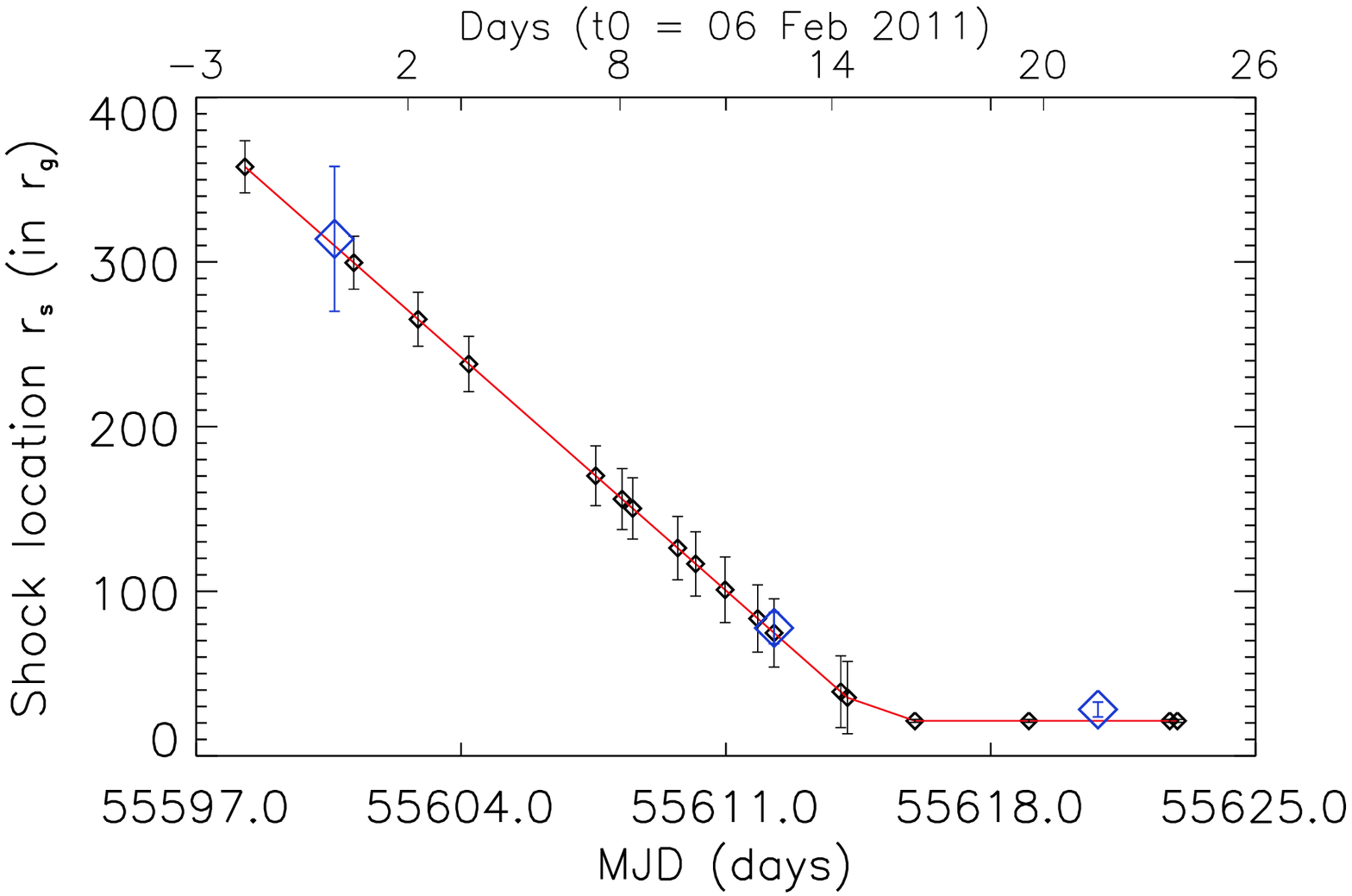}
\caption{Evolution of QPO frequencies with time (days) of the onset-phase of the 2011 outburst of
IGR J17091-3624 is fitted with POS model solution (top panel).
Marked points (with cross) in top panel are {\it Swift XRT} QPOs.
The corresponding size of the oscillating 
region is shown in the bottom panel with diamonds (blue) showing
the `Compton cloud' size as determined from spectral modelling (see \S \ref{ss:spe}). 
The best fit curve (red solid line) of Eqn. 2 to 4 gives an estimate of mass to 
be $14.37^{-0.67}_{+0.71} \Msun$.}
\label{fig:qpd}
\end{figure}

We model the monotonic increase in QPO frequencies with time using the Propagating Oscillatory Shock (POS) 
model (\citealt[]{cha00}, hereafter CM00; \citealt[]{cha08}). This model which comes under the TCAF 
paradigm, explains formation of QPOs due to an oscillating shock front formed in 
the sub-Keplerian component of the flow. 
The shock front starts to oscillate at infall time scales, where the infall time is the 
time taken by material parcel to fall onto the central source from the shock location.
The reasons for shock oscillation can be either resonance oscillations, when the infall time scale
is comparable to the cooling time scale of the Comptonizing post-shock region \citep{mol96}, or
unstable perturbations occurring in the shocked flow due to its viscosity \citep{lee11,das14}.
The frequency of the QPO, which is the rate at which the shock front oscillates \citep{cha00,cha08,cha09,nan12,rad14}
is given as :

\begin{equation}
  \nu(r_s) =  \frac{\nu_{s0}}{2 \pi R r_s \sqrt{r_s - 1.0}},
  \label{eq:qde}
\end{equation}

where, $r_s$ is the location of the drifting shock front, $\nu_{s0}$ is the inverse 
of light-travel time across the black hole, given as $\nu_{s0} = \frac{c}{r_g}$ and $R$ is 
the compression ratio (i.e., ratio of post-shock to pre-shock densities of the flow). In 
Eqn. \ref{eq:qde}, we have introduced an additional $2\pi$ factor in the denominator to ensure that 
the entire axisymmetric shock front region ($2\pi r_s$) is oscillating in-phase and this is 
consistent with the hydrodynamic simulations of \citet{mol96}. The shock front drifts inward 
with a constant velocity $v_0$ giving, 
\begin{eqnarray}
  r_s(t) = r_{s0} \pm v_0(\frac{t - t_0}{r_g}) ;\;\; r_s > r_{s,min}, \\
  r_s(t) = r_{s,min} ;\;\; \mathrm{otherwise},
\label{eq:qrs}
\end{eqnarray}
where $r_{s0}$ is the initial shock location at
$t_0$ ($1^{st}$ day of the outburst). We think that the observed flattening/saturation of QPO frequencies
(see Figure \ref{fig:qpd}) may be due to the fact that the oscillating shock front stops  
drifting further inward.
We account for this in the equation by using a minimum shock location $r_{s,min}$. 

We do the QPO evolution fitting by again using Markwardt's IDL routines$^{\ref{fn:mrk}}$.
The difference between our attempt and previous studies of the POS model
\citep{cha08,cha09,nan12} is that we have used mass as a free parameter. 
Most of the published work before this has dealt with
establishing the POS model for sources of known mass by demonstrating that the QPO
evolution can indeed be explained using the simple parametric expression in Eqn.~\ref{eq:qde}.
In our current implementation, the Schwarzschild radius $r_g$ which is proportional 
to the mass of the source is used in the fitting function (Eqn.~\ref{eq:qde}-\ref{eq:qrs}), 
as a free parameter. Hence mass is calculated by determining $r_g$. 
The other free parameters in the fitting are $r_{s0}$ (the initial shock location) and $r_{s,min}$.
The first day of the outburst ($t_0$) is set from observations at 06 Feb 2011 (MJD 55598.28). 
$R$ is held fixed at 3.0 and $v_0$ at 10 m/s, which are typical values for these parameters.
This is inferred from the fact that the shock front oscillates to give a detectable QPO 
for $R$ between 2.0 and 4.0 \citep{nan12}. The typical inward drift velocity of this front
as obtained from previous studies lies between 5 -- 15 m/s \citep{cha08,cha09,nan12,rad14}. 
We use the mean of these ranges in our fitting and investigate the effects
of variation of $R$ and $v_0$ in the discussion section.

The results of the POS model fitting is shown in Figure \ref{fig:qpd} along with residuals (bottom 
panel of the top figure). The bottom figure shows the variation of the Comptonizing region 
(i.e., the size of the oscillating region) which decreases as the QPO frequency increases with time. 
The dashed dotted vertical line marks the transition of hard-intermediate to soft-intermediate state 
\citep{iye13}, where photon index saturates over
QPO frequency (see Fig. \ref{fig:qpd} and Fig. \ref{fig:gnc}). We obtain the fitted value of the mass to be $\mathrm{M_{bh} = 
14.37^{-0.67}_{+0.71} \Msun}$ with $\chi_{red}^2$ = 1.03 ($\chi^2/dof = 15.43/15$). 

\subsection{Spectral modelling based on TCAF} \label{ss:spe}
The spectral modelling is based on steady state hydrodynamic solutions of the equations governing
the flow of accreting material under the TCAF paradigm. The solution for a set of free parameters 
gives a model spectrum, which
we use to compare and fit the observed spectrum. We have incorporated 
the TCAF model \citep[]{cha95,cha06,mc10} in {\tt XSPEC} as a local model \citep{arn96} 
to achieve this spectral fitting. In this model, 
we consider a sub-Keplerian flow of accretion rate $\dot m_h$ on the top of a Keplerian flow of 
accretion rate $\dot m_d$ at the equatorial plane. Both the accretion rates are measured in units of 
the Eddington accretion rate ($\dot m_{Edd}$). The other free parameters are mass of the central 
object ($M_{bh}$), shock location ($r_s$) and compression ration ($R$). 
A similar implementation in validating the TCAF
model for a few black hole sources has been carried out by \citet{deb14}. But these previous studies are
limited within 3.0 keV -- 30.0 keV whereas in the current work we model the broadband spectrum between 0.5 keV --100 keV.

To implement the local model in {\tt XSPEC}, we use the {\tt HEASoft} facility to create a table model 
by generating the model spectra for a large range of input parameters and saving it
in an array. This table model can be run from {\tt XSPEC} ({\tt XSPEC} User's Guide) to calculate the fitting parameters. 
However, the values of the fit parameters so obtained may not be very
accurate as {\tt XSPEC} does an interpolation to fit the data. 
Hence, we have done the spectral modelling in two steps. First, we use the table model to constrain the 
parameters in reasonably narrow interval and then we run the actual hydrodynamic source code from {\tt XSPEC}
with the parameters previously obtained narrow interval to get the best fit.

During the 2011 outburst, the source evolved from hard to soft state via intermediate states.
We have fitted the broadband spectra (0.5 - 100.0 keV) for all the observations where we could find
simultaneous or quasi-simultaneous broadband observations. The observations
so found lie in 
each of these spectral states: hard state (HS) observed on \texttt{08 Feb}, 
hard intermediate (HIMS) on \texttt{20 Feb}, Intermediate state near transition from
HIMS to SIMS on \texttt{22 Feb}, soft intermediate (SIMS) on \texttt{28 Feb}, 
Intermediate state near transition from SIMS to SS on \texttt{08 Mar}   
and soft state (SS) on \texttt{12 Mar} \citep[see][]{iye13}. We model these six
spectra by using our local TCAF model alongwtih {\it phabs} model of {\tt XSPEC} to account for
the interstellar absorption.
In the model spectrum computation, the soft photons are supplied by the Keplerian disk and are
inverse-Comptonized by hot electrons in the post-shock region. Hence, 
the resultant spectrum is a combination of soft and the hard components and the relative 
normalization is determined by the fraction of photons intercepted by the post-shock region. 
We finally fit the resultant
spectra with observed spectra as shown in Fig. 3. The TCAF model fitted parameters are 
presented in Table \ref{tab:res}. We have kept the compression ratio fixed (R=3)
for all spectral fitting. We see that as the source moves to soft state, 
the shock front moves inward and $\dot m_d$ increases while $\dot m_h$ decreases except for
observations near the state transition boundaries (see \S\ref{s:dis} for details). This 
means that in the soft state, a lot more soft photons come from Keplerian disk and 
there exist lesser number of hot electrons to cool down. 
This is consistent with the physical picture of an accreting black hole 
source \citep{cha95,smi01,smi02,cha06,mc10}.

\begin{figure*}[t]
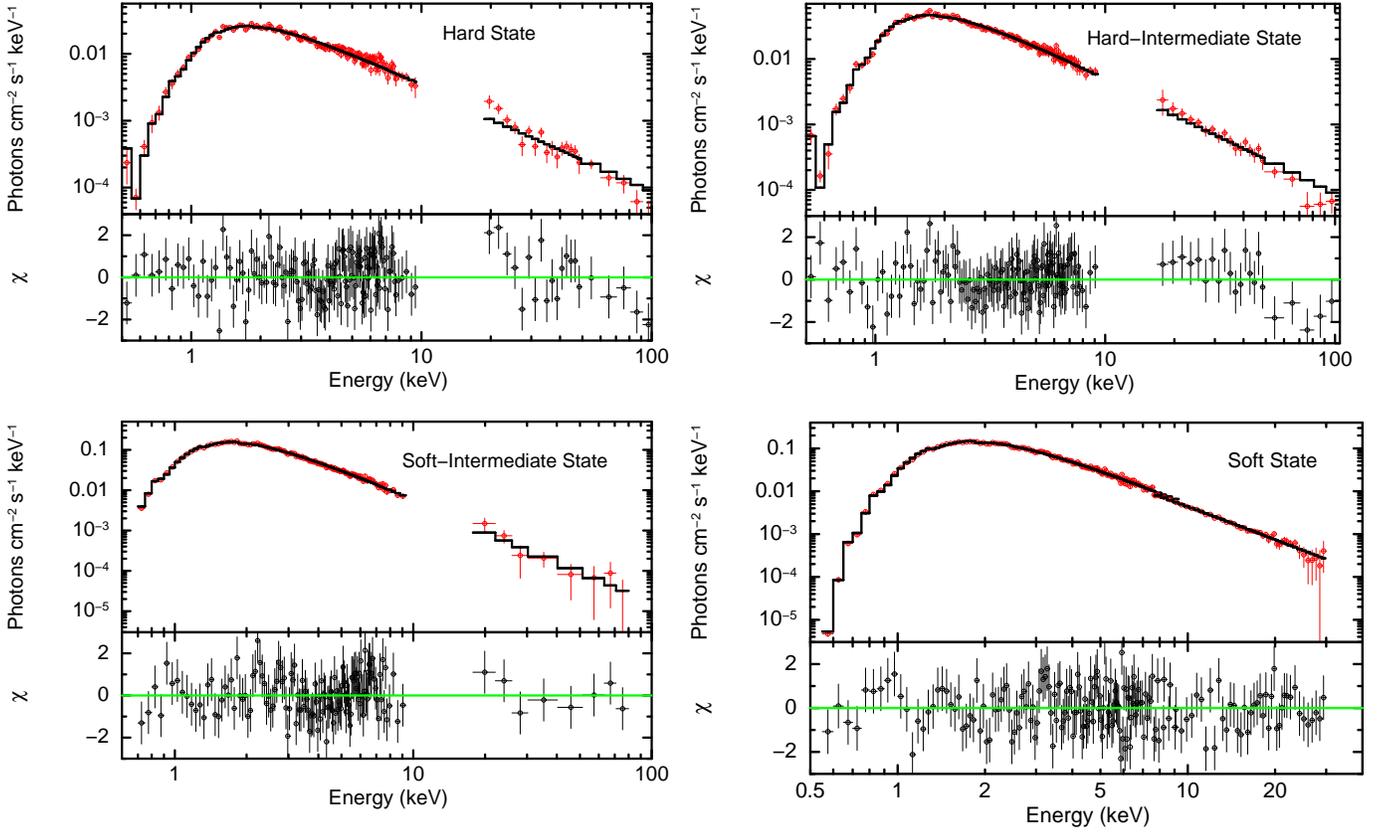

\epsscale{1.0}
\includegraphics[height=0.5\textwidth,angle=270]{f3a.eps}
\includegraphics[height=0.5\textwidth,angle=270]{f3b.eps}
\qquad
\includegraphics[height=0.5\textwidth,angle=270]{f3c.eps}
\includegraphics[height=0.5\textwidth,angle=270]{f3d.eps}
\caption{Broadband energy spectra fitted by TCAF model for different states mentioned on the respective 
figures (residuals are shown in bottom panels). HS, HIMS and SIMS spectra are taken by 
using simultaneous data from {\it Swift XRT} and {\it INTEGRAL IBIS}.
SS spectrum is taken by using simultaneous data from {\it Swift XRT} and {\it RXTE PCA}.
The fit parameters along with the estimated mass are presented in Table \ref{tab:res}.}
\label{fig:tsm}
\end{figure*}

In our spectral fitting, we find that the column 
density ($n_H$) varies between $(0.97-1.7)\times10^{22}cm^{-2}$ while the source transits
from the hard to the soft state. A higher value of $n_H$ in the soft state may be due to enhanced disk
winds \citep[see][]{kin12} or other sources of opacity.
In our present work, we do not assume any distance 
to the source. We calculate the overall normalization parameter (fixed for all
spectral fits) from 
our spectral modelling.
For every data-set, we calculate the normalization constant associated with the best fit and then take an 
average over all normalizations from individual data-sets. Finally, we fit
all the data-sets again with 
this new averaged normalization constant. Also, we do check that the final estimation of mass from individual 
data-sets with fixed overall normalization lies within the range of mass as determined by the individual 
normalization before taking the average. The overall normalization depends on the distance to the 
source and inclination angle of the system.
The reported distance to the source and its inclination angle varies between 10 - 20 kpc \citep{alt11,rao12} 
and 50$^\circ$ - 70$^\circ$ \citep{kin12} respectively.
Hence for an inclination angle in the range (50$^\circ$ - 70$^\circ$), the 
final normalisation constant translates into a distance range (10.6 - 14.6) kpc. For lower 
values of inclination angle, the distance to the source reduces even further. This suggests that 
the source flux is low, not due to the source being located at a very large distance.
The spectral fitting gives a mass estimate as listed in the last column of Table \ref{tab:res}, 
with $\chi_{red}^2$ ($\chi^2$/dof) for different states as 160.57/157 (HS), 
132.85/158 (HIMS), 238.98/269 (HIMS-SIMS), 117.53/138 (SIMS), 341.85/283 (SIMS-SS)
and 163.83/176 (SS) respectively.

\begin{deluxetable*}{l|c|c|c|c}
\tabletypesize{\small}
\tablecaption{Results of the fitting for Mass estimation\label{tab:res}}
\tablewidth{0pt}
\tablehead{
  \colhead{Method} & \multicolumn{3}{c}{Details and Assumptions} & \colhead{Mass estimate}
}
\startdata
(A) $\Gamma$-$\nu$ Correlation & \multicolumn{3}{c|}{See ST07} &  $\mathrm{10.90^{-1.48}_{+1.67}} \Msun$   \\
(B) QPO Evolution (POS model) & \multicolumn{3}{c|}{See CM00, \citet{nan12}} & $14.37 \pm 0.67 \Msun$  \\\\
(C) Spectral Modelling (TCAF) & State & Shock location & {Accretion rate} & {Mass} \\
 & \nodata  & ($r_s$) & ($\dot m_h$, $\dot m_d$)  &  ($\Msun$)\\
 & HS & $314 \pm 44$ & $0.66\pm0.04$, $0.08\pm0.01$  & $12.7 \pm 0.6 $ \\
 & HIMS & $77.3 \pm 9.3$ & $0.58 \pm 0.04$, $0.12 \pm 0.02$  & $10.7 \pm 0.5 $ \\
 & {HIMS-SIMS} & $\mathrm{ 35.1 \pm 11.3}$ & $\mathrm{ 0.16 \pm 0.015}$, $\mathrm{ 0.21 \pm 0.14}$  & $\mathrm{ 11.5 \pm 1.5 }$ \\
 & SIMS & $28.2 \pm 4.5$ &  $0.28 \pm 0.05$, $0.63 \pm 0.08$  & $12.9 \pm 1.1 $ \\
 & { SIMS-SS} & $\mathrm{ 24.0 \pm 11.6}$ &  $\mathrm{ 0.12 \pm 0.02}$, $\mathrm{ 1.09 \pm 0.09}$  & $\mathrm{ 13.1 \pm 1.1}$ \\
 & SS & $19.0 \pm 3.7$ & $0.25\pm0.06$, $0.64\pm0.12$ & $13.1 \pm 1.6 $ \\\hline \\
 {Final Bounded Mass} & \multicolumn{4}{c}{ Joint Likelihood : { 11.8$\Msun$ -- 13.7$\Msun$}}     \\ 
\enddata
\end{deluxetable*}

\section{Discussion and Conclusions}\label{s:dis}

In this paper, we have tried to constrain the mass of IGR J17091-3624 by attempting to explain 
the X-ray spectral and timing observations under the TCAF paradigm. We find that the source
behaviour can be consistently explained if the source mass is $\gtrsim$ 10$\Msun$.
To quantify this further, we first examine the limitations on each of our methods one by one. 
Finally, we attempt to put a single set of limits on the source mass from the different
estimates presented in this paper. The summary of mass estimates by 
{\it Photon Index ($\Gamma$) - QPO frequency ($\nu$) correlation},
{\it QPO frequency ($\nu$) - Time ($day$) evolution} and {\it Spectral modelling based on TCAF}
is given in Table \ref{tab:res}. 

In the $\Gamma$-QPO correlation model, we have noted that our inferred value
of mass depends on the value of `D' parameter. 
In this method, we find that the range of frequency 
over which the transition (bottom left to top right) in the correlation curve
(see Fig. \ref{fig:gnc})  occurs for the reference 
source of our choice, GRO J1655-40 ($\sim$ 0.1 Hz to 15 Hz), is about 
three times the range for IGR J17091-3624 ($\sim$ 0.05 Hz to 5 Hz).
 This would give us an estimate of D=0.33, 
 obtained as a ratio of the frequency range of the two sources, as D controls the 
range of frequencies over which the correlation curve shows the above mentioned transition.
However, we also note that the value of D = 1.0 is used to scale the mass of GRO J16555-40 
to GRS 1915+105 (see ST07 for details). GRS 1915+105 has a similar span of QPO frequencies like IGR J17091-3624. 
To resolve this, we redo the fit for different values of parameter D from 0.33 to 1.0, and note the variation in mass 
as the uncertainty / limitation of this model. We find that the mass 
estimate shifts from a minimum at 9.50$\Msun$ to a maximum of 10.90$\Msun$. The value of
1.4$\Msun$ (10.9$\Msun$ - 9.5$\Msun$) is total uncertainty on the estimated mass using this method.
To use this uncertainty in our final estimate, we take the 1$\sigma$ uncertainty to be one third
of this value (i.e., we take 1.4 $\Msun$ to encompass 3$\sigma$ (or 99\%) of the total uncertainty 
of this method). The scaling law used to obtain the mass from this method is based on the 
model of \citet{tit04} and the empirical relation of ST07.
If this empirical relation and the subsequent linear scaling mis-specify the actual physical
relation involved, then that could introduce another source of uncertainty in this estimate (see 
\citet{yan12} to get an idea of model uncertainties due to mis-specified scaling assumptions). 
We do not explicitly account for this in our paper, as further investigation of this model would require
cross validating the model against multiple black hole systems with a large sample of data (see \citet{kess13}
for an example of uncertainty estimation).

In the QPO evolution model, we see that the the range of values over which R and $v_0$ can potentially vary 
adds a source of uncertainty to our quoted values of mass. As done above, we try to estimate this by 
redoing the QPO evolution fit for different R (from 2.0 to 4.0) and $v_0$ (from 5 m/s to 15 m/s).
We find that this gives an uncertainty in mass of
1.7$\Msun$ from the mean value (i.e., the standard deviation (1$\sigma$) in the final set of mass values is 1.7$\Msun$). 
Secondly, and more importantly, the data does not fit well if $v_0$ is less than 8 m/s or 
greater than 12 m/s. In a recent paper, \citet{mon15} have mentioned that the velocity of 
propagation of the shock front can be calculated from spectral modelling by estimating the shock location. 
However, in the present paper we treat spectral modelling and QPO frequency modelling as two independent 
methods and hence do not use the results from one method in the other method.

The spectral fitting method under TCAF paradigm also has a few limitations as listed below.
Our spectral modelling includes black body and inverse-Comptonization components. These two components are 
good enough to fit the  spectra of IGR J17091-3624. But there are other sources where these two components 
are not sufficient and we need to include components like a Gaussian emission line profile (e.g. GRS 1915+105), 
 and absorption smear edges (e.g. XTE J1859+226). We need to include other physical processes in our model to generate such 
components. The hydrodynamic solutions used to calculate the model spectra are not self-consistent 
transonic solutions. The number of 
fit parameters can be reduced further if we use transonic solutions. 
To account for the shock transition from pre-shock to post shock region, we fix the value of 
$R=3.0$ (as done in \S\ref{ss:qpo}). We find that spectral 
modelling is not very sensitive to $R$ in our case and do not account for it separately. 
Secondly, an uncertainty in overall normalization (fixed for all data-sets) may lead to the uncertainty 
in the estimated mass. Our spectral modelling constrains the system in such a manner, that we address
the spectral shape along with the overall luminosity simultaneously for multiple
data-sets. In principle, we 
do not expect the normalization to change across data-sets.
Hence, we expect this uncertainty to be small enough to not significantly affect the mass estimation.

\begin{figure}[h]
\epsscale{1.1}
\plotone{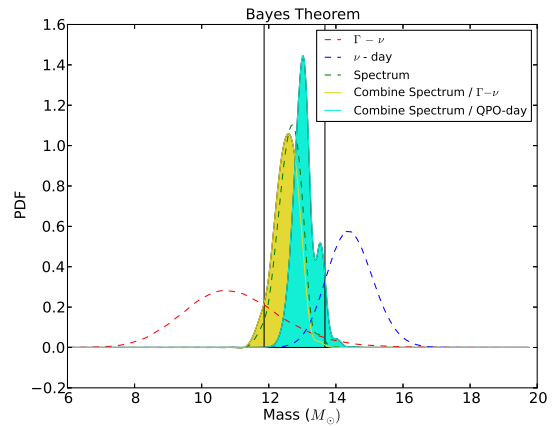}
\caption{Combining the estimates. Figure shows the results from 
the joint likelihood based approach for combining the mass estimates.
See Appendix II and Appendix III for details.}
\label{fig:com}
\end{figure}

Given these three estimates and considering the limitations of the methods, we try to
combine the results to get the overall range of mass of the object.
For combining the estimates, we calculate the probability distribution function (PDF)
of the mass of the central object from the least-squares fit for each method (see Appendix II).
We combine results from the independent methods only. 
Thus, we do not combine the results from 
$\Gamma-\nu$ correlation and from QPO-day variation, since both these methods share the common variable 
$\nu$ (the QPO frequency). Instead we combine results from \S 3.1 with
results from \S 3.3, and results from \S 3.2 with results from \S 3.3 
separately. 
Finally, we quote the lowermost and uppermost values from such a
combination as the final mass bound.
The approach that we take to combine the estimates is based on estimation of 
joint likelihood using the Naive Bayes theorem from independent PDFs (see Appendix III for details).
We present the combined estimate in Figure \ref{fig:com}, where red, blue and green lines are individual
PDFs from \S 3.1, \S 3.2 and \S 3.3 respectively. The combined estimates are represented by shaded regions.
The combined estimate using this approach for \S $3.1 -$ \S 3.3 gives us a mass range of $11.8\Msun - 13.1\Msun$ 
and for \S $3.2 -$ \S 3.3 gives us a mass range of  $12.5\Msun - 13.7\Msun$. This
gives us a limit on the black hole mass of $11.8\Msun - 13.7\Msun $. For a worst case
estimate, we find the 90\% confidence value limits for each estimate from its respective PDF.
The lowermost and uppermost of these limits places the mass between $8.7\Msun$ –- $15.6\Msun$.

Whichever way we combine the estimates, we see that the mass is greater 
than 8.7 $\Msun$. This suggests that IGR J17091 also harbours a black hole with
mass similar to the black hole of GRS 1915+105 \citep[see][for a recent mass estimate]{rei14}. 
The reason for any distinction between these two sources, may then be the 
difference in mass accretion rates. For IGR J17091-3624, we obtain the total mass accretion rate 
to be in the range (0.4 - 1.1) times the Eddington accretion rate (see Table~\ref{tab:res}). 
We obtain near-Eddington accretion rate close to the transition from SIMS to Soft State and 
lowest accretion rate near the transition from HIMS to SIMS \citep{iye13}. We note that the 
non-simultaneous nature of the spectra in these two observations with a gap of $\sim$ 6 hours may 
have contributed to this. This may also be due to many other time dependent nonlinear effects 
(for e.g. flow viscosity \citep{mc10}, or jet activity \citep{rad14}) involved in the system during the
state transitions. Our spectral fitting is done by using a steady state model and does not account
for these time varying phenomena. We have attempted to apply our spectral modelling techniques on 
GRS~1915+105 as well for a relative comparison of flow accretion rate.
While modelling the $\chi$ class spectrum of GRS~1915+105, we find evidence for 
super-Eddington accretion rates ($\sim$ 6.5 times the Eddington accretion rate)
in the system. In future, we would like to do a detailed broadband spectral modelling of different variability 
classes of GRS~1915+105 and IGR J17091-3624 to make a comparative study of the variation in accretion rate 
in both systems. However, it is suggestive to note that the difference in accretion rates, 
could be the reason for the observed difference in X-ray flux. 

If that indeed is the case, then it raises the question of how both sub- and super-Eddington 
accretion rates give rise to similar kinds of multiple variability classes as seen in these objects. 
It is possible that the complex variability classes observed in these sources 
(IGR J17091-3624 \& GRS 1915+105) 
could be due to the interplay between the Keplerian and sub-Keplerian flows in presence 
of outflows/winds in the soft-intermediate state \citep{cha95,cha01,mc10}. 
As seen from Table \ref{tab:res}, the intermediate states
have comparable values of sub-Keplerian ($\dot m_h$) and Keplerian ($\dot m_d$) accretion rates, 
which can be due to the conversion of one type of matter flow into the other type \citep{mc10}. 
Accordingly, a super-Eddington accretion rate may not be a pre-requisite for these 
variability classes to occur. The interplay between the different accreting (Keplerian and sub-Keplerian)
matter with the outflowing matter may itself cause such variabilities. 
It is possible that such an interplay between the
accreting and outflowing streams is currently happening in GRS 1915+105 and that in the past 
this system too might have undergone an evolution from the Hard to the Intermediate 
states to its present phase like other such sources (see \cite{nan12} for GX 339-4 and 
\cite{cap12} for IGR J17091-3624). 

Details of such similarities between the variability classes in GRS 1915+105 and those observed
in other such sources will be explored later and presented elsewhere.
Until we know a concrete explanation for the occurrence of such variability classes, 
concluding on the nature of sources showing such variabilities
is difficult. However, our current work suggests the presence of a high mass stellar
black hole in the binary system IGR J17091-3624, which accretes at sub-Eddington rates and still
shows such variability classes.  

\acknowledgments

This research has made use of data obtained through the High Energy Astrophysics Science Archive 
Research Center (HEASARC) online service, provided by the NASA/Goddard Space Flight Center.
We thank Prof. Belloni and Prof. Markwardt for making their codes (GHATS and MPFIT) available.
We thank Prof. A. R. Rao, TIFR and Dr. Seetha, Space Science Office, ISRO-HQ for 
useful suggestions. We thank Dr. Kim Page of the {\it Swift} team for prompt and continued help in 
the analysis of {\it Swift XRT} data. We thank Dr. Sankar, ISAC, Prof. T. Krishnan 
(retd.) and Prof. M. Delampady, ISI for their help and suggestions with error analysis and statistics.
We thank Dr. Anil Agarwal, GD, SAG; Mr. Vasantha E. DD, CDA  and Dr. S. K. Shivakumar, Director, ISAC 
for encouragement and continuous support to carry out this research. Finally, we would also 
like to thank anonymous referees for their useful suggestions to improve the
manuscript.

\appendix{}

\section{Appendix I : Fitting with errors in two variables}
This gives a brief outline of the error in variable method, and
the technique that we have used here. For further details, please refer to 
\citet[][and references therein]{mac92,pre92}. One way to accommodate errors
in the second variable is to alter the fit statistic used for minimization.
Various formulations are discussed in \citet{mac92}, but we use the one
(denoted here as $\mathcal{O}$) and given as 
\[{ \mathcal{O}^2 = \sum\limits_{i=0}^{N} \frac{(y_i - f(x_i))^2}{var(y_i - f(x_i))}} \; . \]
For estimating confidence intervals on the mass, we do Monte Carlo
runs of the same fitting routine with the observations ($x_i\,,y_i$) 
randomly generated as per their error variance \citep[see][for details]{pre92}.
We then use a Gaussian kernel density estimator to find the probability density function of the
obtained set of mass values. The errors on the final set of parameters are
quoted as 1$\sigma$ (68\%) deviations from their central value.

\section{Appendix II : PDF calculation for the methods mentioned in \S 3.1 - \S 3.3}
\begin{figure}
  \epsscale{0.5}
  \plotone{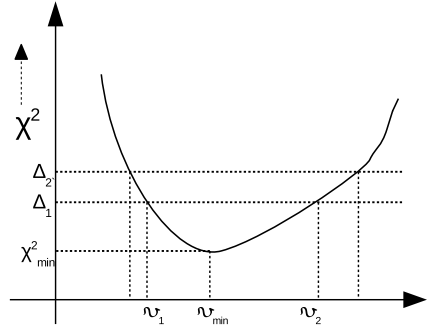}
  \caption{Chi-square variation plot of parameter ($v$) used to obtain the PDF. Plot shows
  different levels of $\chi^2$, namely $\Delta_1$ at which the confidence levels and the confidence
  intervals ($v_1,v_2$) are estimated. The PDF is calculated from these confidence levels.}
  \label{fig:app2}
\end{figure}

This section gives an outline of the methods used to construct a PDF from the
results of the statistical fitting routines. 
\begin{itemize}
  \item Construct PDF for \S 3.1 \\
The PDF is obtained by doing Monte-Carlo simulation runs on the data. 
We simulate multiple data-sets ($\sim 10^5$) from the assumed Gaussian error 
distributions on photon index ($\Gamma$) and QPO frequency ($\nu$) (Appendix I). 
For each simulated data-set we repeat the least squares fitting routine and note
the parameter and statistic value. Finally, we obtain a distribution of 
our parameter of interest (i.e., Mass) by using a Gaussian Kernel Density Estimator.

  \item  Construct PDF for \S 3.2 and \S 3.3 \\
The PDF is obtained from the variation of the fit-statistic (chi-square) about the parameter minima. 
This gives the confidence intervals ($v_1, v_2$) on a single parameter from the least-squares fit as
seen in equation below

\[  C^{v_2}_{v_1} = \int^{v_2}_{v_1} p(v)dv = P(\chi^2 > \Delta_1) .\]

Here, $\chi^2$ denotes the chi-squared distribution with 1 degree of freedom, 
$p(v)$ denotes the PDF of parameter $v$, $C^{v_2}_{v_1}$ denotes the confidence level
and $\Delta_1$ is the level (corresponding 
to $v_1$ and $v_2$) of variation in the $\chi^2$ statistic (see Figure \ref{fig:app2} 
for details). 

These confidence levels are taken to be the representative areas under the 
PDF curve in the given interval ($v_1$ to $v_2$). We find the half area (denoted as $A$) under 
the PDF curve from each side of the interval ($v_1$ or $v_2$) to the minimum value ($v_{min}$)
by

\[ A^{v_{min}}_{v_1} = \frac{v_{min} - v_1}{v_2 - v_1} C^{v_2}_{v_1} \; ; \; A^{v_2}_{v_{min}} = \frac{v_2 - v_{min}}{v_2 - v_1} C^{v_2}_{v_1} .\]

This is applied to $\Delta_1$ which is closest to $v_{min}$. For subsequent $\Delta$ and $v$, we find 
the corresponding half areas ($A$), by cumulatively adding to the previous estimate of half area ($A$).
Thus, we bin the $\Delta$ values in small steps to estimate these half areas for
many points. From these half areas (A), the Cumulative Density Function (CDF)
for each distribution can be found as,

\[F(v) = \int^v_{-\infty}p(v)dv = \{^{A^{v_{min}}_{-\infty} - A^{v}_{v_{min}}\, ,\, v < v_{min}}_{A^{v_{min}}_{-\infty} + A^{v}_{v_{min}}\, ,\, v > v_{min}} .\] 

Once, we obtain the CDF, the PDF can be calculated by taking a derivative of this, which can be written as : 
\[ p(v) = \frac{d}{dv}F(v) .\]

\end{itemize}

\section{ Appendix III : Combining the estimates}
The combination of estimates can be done using the likelihood $P(M_i | X_i)$, which denotes 
the PDF of mass from observations $X_i$ of method $i$ (see Appendix II for details on estimation of the 
PDF). We follow the approach as explained below to find the combined / overall limits of the mass 
from independent PDF only. Thereby, we combine the estimate in \S 3.1 with \S 3.3 and \S 3.2 with \S 3.3 separately.

\begin{itemize}
  \item  Joint Likelihood estimation using the Naive Bayes theorem \\
The joint likelihood of the three methods for BH mass estimation can be obtained using 
the simple procedure known as the `naive Bayes' algorithm where 
\[ P(M) =  K \prod P(M_i | X_i) .\]
The Naive Bayes approach has been used commonly for multivariate
classification of objects like stars in catalogs \citep[see][]{bro13}. 
We use it here, for parameter confidence level estimation. 
Here, $P(M)$ denotes the combined / final PDF of mass and $K$ is the normalization constant to get
unit area PDF. This method, when applied to Gaussian distributions gives exactly 
the weighted average estimate, where the mean of the joint PDF will give the 
weighted average and the standard deviation of the joint PDF will give the 
error on the weighted average. Thus, in this method the width of the final PDF is 
always less than that of each individual PDF. However, this method can only be applied to PDFs which
are independent to each other. In classification schemes, usage of the
Naive Bayes approach has shown to give acceptable results even if independence is violated,
as is shown in \citet{dom97}. In our case, if dependence among our mass estimation methods exists
and is not accounted for, leads to underestimation of the width of the final PDF. 
To overcome this, we do not combine the PDFs from the methods which may have some level of
dependence. Another limitation of this method is that it gives an offset / biased PDF, if any
of the individual PDFs have unaccounted biases / systematics. To calculate
the joint PDF we perform a multiplication of the individual PDFs for every value of mass 
in the range from $6\Msun$ to $20\Msun$.
\end{itemize}

Once we obtain the final PDF $P(M)$, we quote the confidence interval of
the mass by finding the limits beyond which the PDF encompasses 5\% of the distribution
on each side, i.e., we compute $M_1$ and $M_2$ to give $P( M < M_1) = P(M > M_2) = 0.05$.
Thus, we find the 90\% confidence interval on the mass. We note that this interval
may extend to a smaller level of confidence than 90\% due to an increase in the
width of the combined PDF in case of unaccounted dependence between the individual PDFs.

\end{document}